\documentstyle[12pt,epsfig]{article}
\newlength{\dinwidth}
\newlength{\dinmargin}
\newcommand{\ba}{\begin{array}}
\newcommand{\ea}{\end{array}}
\newcommand{\be}{\begin{equation}}
\newcommand{\ee}{\end{equation}}
\newcommand{\bea}{\begin{eqnarray}}
\newcommand{\eea}{\end{eqnarray}}

\def\bee{\begin{eqnarray}}
\def\eee{\end{eqnarray}}
\def\be{\begin{equation}}
\def\ee{\end{equation}}

\begin{document}
\thispagestyle{empty}
\addtocounter{page}{-1}
\begin{flushright}
SNUTP 98-089\\
{\tt hep-th/9807241}\\
Expanded Version
\end{flushright}
\vspace*{1.3cm}
\centerline{\Large \bf Holographic Principle}
\vskip0.3cm
\centerline{\Large \bf and}
\vskip0.3cm
\centerline{\Large \bf Topology Change in String Theory~\footnote{
Work supported in part by KOSEF Interdisciplinary Research Grant 
98-07-02-07-01-5, Ministry of Education Grant 98-015-D00054, and 
The Korea Foundation for Advanced Studies Faculty Fellowship.}}
\vspace*{1.7cm} \centerline{\large \bf Soo-Jong Rey}
\vspace*{0.8cm}
\centerline{\large\it Physics Department, Seoul National University,
Seoul 151-742 KOREA}
\vspace*{1.5cm}
\centerline{\Large\bf abstract}
\vspace*{0.5cm}
D-instantons of Type IIB string theory are Ramond-Ramond counterpart
of Giddings-Strominger wormholes connecting two asymptotic regions of
spacetime. Such wormholes, according to Coleman, might lead to spacetime 
topology change, third-quantized baby universes and probabilistic 
determination of fundamental coupling parameters. Utilizing holographic
correspondence
between $AdS_5\times M_5$ Type IIB supergravity and $d=4$ super Yang-Mills
theory, we point out that topology change and sum over topologies not only
take place in string theory but also are required for consistency with 
holography. Nevertheless, we argue that the effects of D-instanton 
wormholes remain completely deterministic, in sharp contrast to Coleman's 
proposal.
\vskip1.5cm
\leftline{\tt Keywords: \rm Holography, Topology Change, String Theory}
\baselineskip=18pt
\newpage

\setcounter{equation}{0}
One of the most vexing problems in quantum gravity has been the issue of 
spacetime topology change: whether topology of spacetime can fluctuate and,
if so, whether spacetime of different topologies should be summed
over. Any positive answer to this question would bear profound 
implications to the fundamental interactions of Nature. For example, in  
Coleman's scenario \cite{coleman} of baby universes \cite{baby}, fluctuation of
spacetime topology induce third quantization of the Universe and effective 
lon-local interactions thereof~\footnote{
Prior to Coleman's proposal, topology 
fluctuation and possible loss of quantum coherence have been raised
by Hawking \cite{hawking}, Lavrelashvili, Rubakov and Tinyakov \cite{rubakov},
and Giddings and Strominger \cite{giddingsstrominger1}.}. 
As a result, the cosmological constant is no
longer a calculable, deterministic parameter but is turned 
into a quantity of probabilistic distribution sharply peaked
around zero. Likewise, all other physical parameters are renormalized into 
quantities of probabilistic distribution~\footnote{Similar conclusions have been
drawn by Giddings and Strominger~\cite{giddingsstrominger2}.}. 
While the original Coleman's
scenario was based on a Universe with positive cosmological constant, the
topological fluctuation ought to persist to Universes with zero or negative
cosmological constant. For instance, within saddle point approximation of
Euclidean quantum gravity, Carlip \cite{carlip} has argued that topology of
the spacetime with negative cosmological constant is dominated by those 
with extremely complicated fundamental groups only yet having exponentially 
growing density of topologies. 
 
String theory is the only known consistent theory of quantum gravity. As 
such, it would be desirable to address the issue of spacetime topology
change within string theory and draw some definitive answers. 
The questions one would like to understand are: does spacetime topology 
change? what is the rule of summing over topologies? do topological 
fluctuation lead to third
quantization of the Universe? are fundamental physical parameters distributed
probabilistically rather than deterministic quantites? 

In this
paper, utilizing {\sl holographic principle} \cite{holography, susskindwitten}
that underlies string theory and 
consequent correspondence between anti-de Sitter string vacua and boundary
conformal field theories \cite{maldacena, refined}, 
we argue that some answers (albeit being only qualitative and
technically limited) to the above questions can be obtained. In doing so, 
for the sake of concrete setup, we will
restrict the entire foregoing discussions to semi-classical limit of 
Type IIB string theory on $AdS_5 \times S_5$. We will find that the holographic
 correspondence of the $AdS_5$ Type IIB string vacuum to large $N$ limit of 
${\cal N}=4$ super Yang-Mills theory~\cite{maldacena} implies that 
(1) topology change does occur, (2) different 
topologies should be summed over, yet (3) only `tree-level' processes of third
quantization take place, and (4) no probabilistic distribution of physical
coupling parameters is induced.

The argument is extremely simple. Among the quantum solitons in Type IIB 
string theory is the D-instanton \cite{dinstanton} carrying Ramond-Ramond 
axion charge. As observed in \cite{dinstanton} already, the D-instantons
are nothing but Ramond-Ramond counterpart of wormholes (Euclidean gravitational
instantons with axion charge) considered
by Giddings and Strominger \cite{
giddingsstrominger3} previously~\footnote{To be more precise, there is a slight
distinction between the Gidding-Strominger wormholes and D-instantons. The 
wormhols considered by Giddings and Strominger are necessarily 
non-extremal NS-instantons whereas the D-instantons we will consider 
throughout this paper are extremal.}. In string frame, these D-instantons are 
Einstein-Rosen bridges connecting two asymptotic Euclidean spacetime, through
which dilaton and RR axion field vary monotonically.  

Consider a D-instanton in $AdS_5$ Type IIB string background. The configuration
has been studied recently \cite{chuhowu, koganluzon} and we repeat some aspects
needed for foregoing discussions. After Wick rotation to Euclidean spacetime, 
covariant Type IIB equations of motion read
\bee
R_{MN} - {1 \over 3!} ({F_5}^2)_{MN} &=& {1 \over 2} \Big( \partial_M \phi 
\partial_N \phi - e^{2 \phi} \partial_M a \partial_N a)
\nonumber \\
\nabla^M \Big(e^{2 \phi} \nabla_M a \Big) &=& 0 \nonumber \\
\nabla^2 \phi + e^{2 \phi} (\partial a)^2 &=& 0.
\eee
Here, $F_5$ denote self-dual 5-form field strength and
$\phi, a$ are Type IIB dilaton and (Wick rotated) Ramond-Ramond axion fields.
The complex field $\tau = (a / 2 \pi) + i e^{- \phi}$ transforms as a
doublet under Type IIB $SL(2, {\bf Z})$ duality. 
The $AdS_5 \times S_5$ corresponds to the Type IIB background with flux of 5-form field strength $\oint_{S_5} F_5 = N$ while keeping 
the dilaton and the Ramond-Ramond axion fields constant:
\bee
ds^2_{\rm AdS_5 \times S^5 } &=& \alpha' \left[ {U^2 \over R^2} 
d {\bf x}^2 + {R^2 \over U^2} dU^2 + 
R^2 d \Omega_5^2 \right],
\qquad \qquad
(R^2 = \sqrt{4 \pi g_{\rm st} N} \alpha' )
\nonumber \\
F_5 &=& d \left({R^4 U^4 \over 4}\right) \wedge d y^0 \wedge \cdots 
\wedge d y^3.
\label{adsbackground}
\eee
Here, $({\bf x}, U)$ and $\Omega_5$ parametrize Euclidean $AdS_5$ and $S_5$,
respectively, and $R$ denotes the radius of curvature of both $AdS_5$ and 
$S_5$. As the background is non-dilatonic (i.e. dilaton is constant), 
the spacetime geometry is the same for both string and Einstein frames. 
In the low-energy and strong `t Hooft coupling limit,  
$\alpha' \rightarrow 0$ and $g_{\rm st} N \gg 1$,  the string theory is well 
approximated by $AdS_5$ supergravity. 
 
The D-instanton is a ${1 \over 2}$-BPS configuration satisfying the
first order relation: 
\be
d \tau_2 = \pm d \tau_1; \qquad \qquad
 d \Big(e^{-\phi} - e^{-\phi_\infty} \Big) = \pm d 
\Big( a - a_\infty \Big).
\ee
Clearly, contribution of the D-instanton to the energy-momentum tensor 
vanishes identically
and hence the spacetime geometry
 in Einstein frame remains $AdS_5 \times S_5$.
The D-instanton configuration is then specified uniquely by a 
harmonic function $H_{-1} = ( e^{\phi} - e^{\phi_\infty})$, which solves 
the Laplace equation on $AdS_5 \times S^5$. For homogeneous configuration
on $S^5$, the solution is given by \cite{chuhowu} :
\be
H_{-1} (U , {\bf x}; U_0, {\bf x}_0) = \left( e^\phi - e^{\phi_\infty} \right)
= {3 N_{-1} \over \pi^4} \alpha'^4 
{U^4 U_0^4\over \left[ \left( U^{-1} - U_0^{-1}\right)^2  + 
\left({\bf x} - {\bf x}_0 \right)^2 \right]^4}.
\ee
The instanton is centered at $(U_0, {\bf x}_0)$ in $AdS_5$~ \footnote{ 
D-instanton configuration centered at generic point in $AdS_5$ is found by 
\cite{bianchi} (See also \cite{chuhowu}). It is also important to keep
the asymptotic values of the dilaton and the Ramond-Ramond axion fields to 
those 
of gauge coupling and $\theta$-parameter of the super Yang-Mills theory at
the boundary}. 
Consequently, the dilaton and the axion fields are no longer constant but
varies over the spacetime. Hence, it is more natural to describe the
spacetime geometry in the string frame:
\be
ds^2_{\rm string} = \alpha' \sqrt{H_{-1}} \left[
{U^2 \over R^2} d {\bf x}^2 + {R^2 \over U^2} dU^2 + R^2 d \Omega_5^2 \right].
\ee
Because of the harmonic function factor in front, it is straightforward 
to see that the spacetime in string
frame takes precisely the (Euclidean form of) Einstein-Rosen bridge 
connecting
two asymptotic $AdS_5 \times S_5$ regions at $U \sim 0, \infty$.  
The D-instanton carries RR-axion charge $N_{-1}$, which is quantized 
according to Dirac-Schwinger condition in units of inverse charges of
Type IIB D7-branes.

We have seen that, in many respects, 
the D-instantons in $AdS_5 \times S_5$ background
behaves as Giddings-Strominger type Euclidean wormholes.
Being so, one might consider semi-classical limit of Type IIB string theory
in $AdS_5$ background and ask whether Coleman's scenario of baby universes
can be realized. Suppose we place dilute gas of D-instantons on the 
$AdS_5$ background (which will be referred as `base universe'). 
The effects of
D-instanton wormholes are then computed in the standard way. The Type IIB 
supergravity fields are written as background fields, slowly varying on
the string scale, and the D-instanton wormholes are placed in them. The
integration over all possible D-instantons would then produce an 
effective interaction of the background Type IIB supergravity fields on 
the base $AdS_5$ universe. 

There are several possible combinatorics of D-instanton wormhole 
configurations. One possible configuration 
is to branch off a baby $AdS_5$
universe from the base $AdS_5$ one. Another possibility is that, if at
all allowed, the
two ends of D-instanton wormhole are attached both to the base $AdS_5$
universe. From the third quantization point of view, 
the first possibility corresponds to a tree-level process, while the
second one is a loop-process.

Taking into account of effective interactions produced by both 
combinatorics of D-instanton configurations, the resulting low-energy 
Type IIB supergravity action on the base $AdS_5$ universe takes the form:
\be
S_{\rm eff} = S_0 + \sum_I \int_{AdS_5 \times S_5} d^5 x C_I {\cal O}_I (x) 
+ \sum_{I, J} {\int \int}_{AdS_5 \times S_5} d^5 x d^5 y \,\,
C_{IJ} {\cal O}_I (x) {\cal O}_J (y) + \cdots.
\ee
Here, ${\cal O}_I$ denote the supergravity local operators and 
$C_I, C_{IJ}$ are numerical coefficients that encode the D-instanton
wormhole combinatorics: the $C_I$'s are due to tree-level processes
and the $C_{IJ}$'s are due to loop-level processes respectively. 

As in the Coleman's scenario \cite{coleman} of baby universes, 
an important point is that the coefficients $C_{IJ}$'s are 
independent of locations of operator insertion on $AdS_5$. Being
induced by the second possibility mentioned above (i.e. loop-level
processes), this class of effective interactions would then
give rise to non-local effects to the low-energy Type IIB supergravity.      
Should the non-local, loop-level processes are permitted, then physics of
semi-classical Type IIB string theory
 is described by a probabilistic distribution
of physical coupling parameters. 
Introducing coherent state parameter $\alpha_I$'s, one can re-express 
\be
\exp \Big( -S_{\rm eff} \, \Big) = \Big\langle \exp ( - S_{\rm IIB} [\alpha] ) 
\Big\rangle_\alpha \, ,
\ee
where
\be 
\Big\langle \hat A \Big\rangle_\alpha \equiv 
\int [\prod_I d \alpha_I] \, \hat A \,
\exp \Big( - {1 \over 2} K_{IJ} \alpha_I \alpha_J \Big), \qquad \qquad 
(K_{IJ} C_{JK} = \delta_{IK}),
\ee
and
\be
S_{\rm IIB}[\alpha] = S_0 + \sum_I (C_I + \alpha_I)
\int_{AdS_5 \times S_5} d^5 x \, {\cal O}(x).
\label{resultaction}
\ee 
In fact, from the third quantization point of
view, integration over the coherent state
$\alpha$-parameters corresponds to taking into account of loop processes.
In the resulting low-energy supergravity action Eq.(\ref{resultaction}),
the second term 
clearly indicates that physical coupling parameters are not
deterministic but take probabilistic distributions with respect to the
third-quantization $\alpha_I$-parameters. While string theory itself
does not have any adjustable parameters, third quantization of the
universe ought to lead effectively to infinitely many parameters that
would only be determined by probability distribution. To many, this 
is quite an alarming allegation (See, for example, \cite{schwarz}). 

We are thus led to ask back the question: are such non-local, 
loop-processes indeed possible in string theory? \footnote{which we have tacitly assumed to be so 
in deriving Eq.(8).}
 We now would like
to argue that holographic principle does not permit them at all. 

Recently, 
Banks and Green \cite{banksgreen} and Witten \cite{baryon} have identified
that the D-instantons located in the $AdS_5$ space as, once 
holographically projected to the boundary,  
BPS instantons of the $d=4, {\cal N} = 4$ 
super Yang-Mills theory. Being conformally invariant, the super 
Yang-Mills theory does not pose any strong infrared fluctuations. 
As such, at all energy scales, the Yang-Mills instantons are well
defined BPS configurations with quantized topological charges. 
The coordinates of D-instantons on $AdS_5$
are holographically mapped to collective coordinates to the position
and the size of the Yang-Mills instanton at the boundary.
Integration over the position $d^5 x$ of a D-instanton in the bulk of 
$AdS_5$ 
matches precisely to the conformally invariant collective coordinate
measure $U^5_0 dU^{-1}_0 d^4 {\bf x}_0$ of the Yang-Mills instanton (See
Eq.(4))~\footnote{Collective
coordinates associated with the global gauge degrees of freedom are 
expected to get lifted at strong `t Hooft coupling, as $SU_R(4)$ R-symmetry is non-anomalous and index theorem does not protect these zero modes
from being lifted. See the discussion of \cite{banksgreen}.}.
Furthermore, as has been shown explicitly \cite{banksgreen, bianchi},
the tree-level branching-off processes of dilute D-instantons yielding
${\cal R}^4$ interactions (and others related by supersymemtry, for
example, sixteen fermion interactions) is precisely mapped to the
Yang-Mills instanton induced correlators at the boundary.

Using the above holographic relation between D-instantons on $AdS_5$ and 
Yang-Mills instantons at the boundary and our knowledge of instanton
physics in conformally invariant super Yang-Mills theory, we will now be
able to argue that non-local, third quantization loop processes are 
impossible.
Suppose, for the moment, that such processes are present.
According to the holographic principle \cite{holography, susskindwitten}, 
we should identify the effective Type IIB supergravity on the bulk of 
$AdS_5$ with the generating functional of boundary 
conformal field theory. Generic supergravity fields $\psi(U, {\bf x})$
are completely fixed by their boundary values $\psi_\infty ({\bf x})$. 
The nature of this Dirichlet problem should equally work even if the
wormhole-induced effective interactions Eq.(8) are considered. 
Hence, the relation between bulk and boundary correlators would now be
generalized to:
\be
\exp^{-S_{\rm eff}} = \Big\langle \exp^{-S_{\rm IIB}[\psi; \alpha])} 
\Big\rangle_\alpha
= \Big\langle \exp^{- \int d^4 y \psi_\infty \hat{\cal O}_{\rm CFT}} 
\Big\rangle_{\rm instanton}.
\label{correspondence}
\ee
On the right-hand side, the correlators in ${\cal N}=4$ super Yang-Mills 
theory are calculated in the background of Yang-Mills instantons. 

The correspondence Eq.(\ref{correspondence}) poses severe consistency 
problem. Consider correlators of boundary conformal field theory, viz. 
${\cal N}=4$ super Yang-Mills theory. Calculated from
the right-hand side of Eq.(\ref{correspondence}), super Yang-Mills
correlators in the instanton background can be evaluated
unambiguously from the standard instanton physics \cite{banksgreen, 
bianchi} and hence are completely deterministic. On the other hand,
the correlators calculated from the supergravity side do depend
explicitly on the third quantization $\alpha$-parameters and
thus take probabilistic distribution. 
As there simply is no room for probabilistic
$\alpha$-parameters in the ${\cal N}=4$ super Yang-Mills theory, should
the holographic principle remain valid, the 
$\alpha$-parameter dependence somehow out to be suppressed  
in the effective Type IIB supergravity on
$AdS_5$. Tracing back to the origin of the $\alpha$-parameter dependence
in Eqs.(7-9), this is possible only if the integration over the
$\alpha$-parameters are not present. This then follows if the
D-instanton wormholes are not allowed to attach their both necks to the
base $AdS_5$ universe and consequently induce non-local, loop processes.

Wonderfully, $AdS_5$ background achieves this in a rather
transparent way. Recall that the $AdS_5 \times S_5$ space is 
non-dilatonic, viz. the dilaton and the Ramond-Ramond axion fields are frozen 
to constant values throughout the entire $AdS_5$ space.   
The D-instantons, once inserted to the $AdS_5$ space, would only
induce local deformations of these fields around the instanton neck. 
On the other hand, in Eq.(3), we have noted that the dilaton
and the Ramond-Ramond axion fields grows indefinetly at the center, at which it
opens up another asymptotic $AdS_5$ space (baby $AdS_5$ universe). 
Therefore, once one of the D-instanton neck
is attached to the non-dilatonic base $AdS_5$ background, there is no 
way that the
other end of the D-instanton wormhole can join back to the same 
base $AdS_5$ background while maintaining smooth 
local deformations of dilaton and Ramond-Ramond axion fields. 
Even if one tries to invoke $SL(2,{\bf Z})$ quotient of the dilaton and 
Ramond-Ramond axion fields, it is easy to see that smooth 
interpolation of these fields 
would not be possible in general~\footnote{Incidentally, since the
argument relies solely on the fact that dilaton and Ramond-Ramond 
axion fields are
constant throughout the anti-de Sitter space, 
the same conclusion can be drawn for other cases of 
AdS/CFT correspondence such as $AdS_3 \times S_3$ associated with 
near-horizon geometry of D1-D5 branes.} . 
  
What about the tree-level process of D-instanton wormholes, for which
only one side of the wormhole is attached to the base $AdS_5$ space and
the other end branches off a baby $AdS_5$ universe? 
As the Yang-Mills instantons in the boundary conformal field theory is 
realized in the AdS/CFT correspondence by D-instantons on $AdS_5$
\cite{banksgreen, baryon}, the D-instanton wormholes that
are attached to the base $AdS_5$ universe on their one ends (but not both)
should be allowed and, in fact, are the only possible combinatorics. 
Once holographically projected to the boundary, each branch-off D-instanton wormhole represents a single Yang-Mills
instanton. From these observations, viz. existence of well-defined 
Yang-Mills instantons and unambiguous calculability of their effects 
within super Yang-Mills theory, we can draw some answers
to the question we have allued at the beginning provided we combine 
them with the holographic principle. First, since 
instanton-induced processes are present at the boundary conformal field
theory, topology change induced
by D-instantons should be permitted in $AdS_5$ vacua of Type IIB string
theory. In other words, there are changes of spacetime topology
in string theory.

Second, the cluster decomposition 
property of the Yang-Mills instantons at the boundary conformal field
theory requires that arbitrary numbers of D-instanton wormholes on $AdS_5$
should be considered to be consistent with holographic principle. This
then implies that universes of different topology (viz. different numbers
of D-instanton wormholes on $AdS_5$) should be summed over. The rule for
summing over the topologies on $AdS_5$ string vacua 
should not be arbitrary but rather be fixed
precisely such that, once holographically projected to the boundary, the
instanton-induced processes in super Yang-Mills theory are reproduced.

Third, since dilaton and axion fields are fixed at the boundary and 
since Yang-Mills instantons of different sizes are completely 
independent, only one side of the D-instanton wormholes can be attached
to the base $AdS_5$ universe. The other ends open up baby universes with
$AdS_5$ geometries which asymptotically approaches a flat spacetime in string
frame. On these `baby' $AdS_5$ universes, the dilaton and axion fields are 
divergent. As was argued above, the baby $AdS_5$ universes bear no effect 
to the base $AdS_5$ space. This then implies that only tree-level processes 
of the third quantization take place and hence third quantization itself is
unnecessary. 

Fourth, the Yang-Mills instantons induce completely deterministic
correlators at the boundary. Holography principle then implies that
the corresponding branching-off processes of the D-instantons in the
$AdS_5$ bulk are also deterministic. From the third quantization point of 
view, since only tree-level processes are permitted, integration over 
$\alpha$-parameters is unnecessary. Henceforth, in Type IIB string
theory, the topology change of spacetime is completely specified by 
the {\sl calculable} coupling parameters ($C_I$'s in Eq.(6)) and 
local operators (${\cal O}_I(x)$ in Eq.(6)) in the effective supergravity 
theory.   

The resulting topology change processes in string theory 
would then look like schematically as in Figure 1. 

\begin{figure}[htb]
\vspace{0.5cm}
\centerline{
\epsfig{file=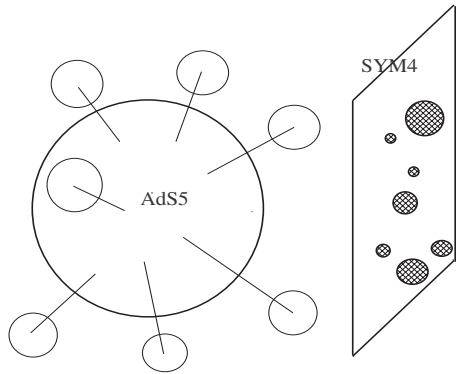, height=2.5in, width=3.5in}
}
\vspace{0.5cm}
\caption{Topology change induced by D-instanton wormholes on $AdS_5$ 
string vacua. Upon holographic projection, they reproduce instanton
induced effects in ${\cal N}=4$ super Yang-Mills theory. Note that the
D-instantons should be interpreted as configurations in the Higgs branch.}
\vspace{1.5 cm}
\end{figure}

To summarize, we have shown that holographic principle draws an interesting 
picture to the issue of topology change in string theory. Not only 
topology change takes place but also should they be summed up. The way
topology is summed up, however, is not arbitrary but should be 
deterministic so that, upon holographic projection to the boundary, 
the standard instanton effects in ${\cal N}=4$ super Yang-Mills theory
are to be reproduced.
This forbids any extra collective coordinates such as Coleman's 
$\alpha$-parameter and hence no probabilistic interpretation either. 
As the argument does not rely on the energy scale of the super Yang-Mills
theory, the conclusion should remain valid in the flat space limit of 
the anti-de Sitter space.
Certainly, most immediate question is whether similar argument can be
applied to situations where the cosmological constant vanishes exactly
as, for example, in M(atrix) theory or take a positive value (cosmological
vacua).  

I would like to thank T. Banks and G. Gibbons for correspondences, 
and to I. Klebanov, N. Seiberg, L. Susskind and E. Witten for discussions.


\begin{thebibliography}{99}

\bibitem{coleman} S. Coleman, {\sl Why There is Nothing Rather Than
Something: A Theory of The Cosmological Constant},
Nucl. Phys. {\bf B310} (1988) 643.

\bibitem{baby} For excellent reviews on baby universe,
see, for example,
A. Strominger, {\sl Baby Universes}, in the proceedings of TASI '88,
pp. 315 - 392 (World Scientific Co. 1988, Singapore);\\
S.W. Hawking, {\sl Baby Universes}, in the proceedings of Friedmann
Centenary Volume, pp. 81 - 92 (World Scientific Co. 1988, Singapore);\\
S. Giddings, {\sl Wormholes, The Conformal Factor and The Cosmological
Constant}, in the proceedings of International Colloquium on Modern
Quantum Field Theory, pp. 518 - 544 
(World Scientific Co. 1990, Singapore);\\
T. Banks, {\sl Report on Progress in Wormholes},
Physicalia {\bf 12} (1990) 19.


\bibitem{hawking} S. Hawking, {\sl Quantum Coherence Down the Wormhole}, 
Phys. Lett. {\bf 195B} (1987) 337.

\bibitem{rubakov} G.V. Lavrelashvili, V.A. Rubakov and P.G. Tinyakov,
{\sl Particle Creation and Destruction of Quantum Coherence by Topological
Change},
JETP Lett. {\bf 46} (1987) 167.

\bibitem{giddingsstrominger1} 
S.B. Giddings and A. Strominger, {\sl Axion Induced Topology Change in 
Quantum Gravity and String Theory}, Nucl. Phys. {\bf B306} (1988) 890.

\bibitem{giddingsstrominger2}
S.B. Giddings and A. Strominger, {\sl Loss of Incoherence and Determination
of Coupling Constants in Quantum Gravity}, Nucl. Phys. {\bf B307} (1988) 
854.

\bibitem{carlip} S. Carlip, Phys. Rev. Lett. {\bf 79} (1997) 4071.

\bibitem{holography} 
G. 't Hooft, {\sl Dimensional Reduction in Quantum Gravity}, in `Salam
Fest' (World Scientific Co. Singapore, 1993);\\
L. Susskind, {\sl The World as a Hologram}, J. Math. Phys. {\bf 36} 
(1995) 6377.

\bibitem{susskindwitten} L. Susskind and E. Witten, 
{\sl The Holographic Bound in Anti-de Sitter Space}, {\tt hep-th/9805114}.

\bibitem{maldacena} J. Maldacena, 
Adv. Theor. Math. Phys. {\bf 2} (1988) 231, {\tt hep-th/9711200}.

\bibitem{refined} S.S. Gubser, I.R. Klebanov and A.M. Polyakov,
Phys. Lett. {\bf B428} (1998) 105,
{\tt hep-th/9802109};\\
E. Witten, Adv. Theor. Math. Phys. {\bf 2} (1998) 253,
{\tt hep-th/9802150} .

\bibitem{dinstanton} G.W. Gibbons, M.B. Green and M.J. Perry,
{\sl Instantons and Seven-Branes in Type IIB Superstring Theory},
Phys. Lett. {\bf 370B} (1996) 37..

\bibitem{giddingsstrominger3}
S.B. Giddings and A. Strominger, 
{\sl String Wormholes}, Phys. Lett. {\bf B230} (1989) 46.

\bibitem{chuhowu} C.-S. Chu, P.-M. Ho and Y.-Y. Wu, 
Nucl. Phys. {\bf B541} (1999) 179, {\tt hep-th/9806103}.

\bibitem{koganluzon} I.I. Kogan and G. Luz\'on, 
Nucl. Phys. {\bf B539} (1999) 121, {\tt hep-th/9805112}.


\bibitem{schwarz} J.H. Schwarz, {\sl Can String Theory Overcome
Deep Problems in Quantum Gravity?}, Phys. Lett. {\bf 272B} (1991) 239;
\\
L. Susskind, private communication (June, 1998).

\bibitem{banksgreen} T. Banks and M.B. Green, 
J. High-Energy Phys. {\bf 9805} (1998) 002, {\tt hep-th/9804170}.

\bibitem{baryon} E. Witten, 
J. High-Energy Phys. {\bf 9807} (1998) 006, {\tt hep-th/9805112}.

\bibitem{bianchi} M. Bianchi, M.B. Green, S. Kovacs and G. Rossi,
J. High-Energy Phys. {\bf 9808} (1998) 013, {\tt hep-th/9807033}.

\end{thebibliography}
\end{document}